# Whispering-Gallery and Luneburg-Lens Effects in a Beam-Fed Circularly-Layered Dielectric Cylinder

Artem V. Boriskin, *Student Member, IEEE,* and Alexander I. Nosich, *Senior Member, IEEE*

***Abstract*—The Whispering-Gallery (WG) mode excitation and Luneburg Lens (LL) effect are studied for a lossy circularly-layered dielectric cylinder illuminated by a beam field. The latter is simulated by the Complex Source-Point (CSP) beam. Exact series solution to the wave scattering problem is used to obtain the far-field patterns and directivity. The WG mode effect is shown to undermine the LL performance.**

***Index Terms*—layered dielectric cylinder, Complex Source-Point beam, Whispering-Gallery modes, Luneburg Lens.**

## I. Introduction

The aim of the paper is to study the effects accompanying a beam transmission through and scattering by a circularly-layered dielectric cylinder. We imply here the excitation of the WG modes and the LL focusing effect. The WG modes have the fields that are confined at the outer boundary of the cylinder [1]. WG modes find applications in optical and microwave resonators, band-stop filters, detectors, etc. [2]. Dielectric lenses are used as focusing devices for a laser beam at the input to an optical fiber. LL is a remarkable special case of lens shaped as inhomogeneous dielectric sphere [3] or circular cylinder. In the microwave range, lenses are candidate antennas for the wide-band multi-beam and mechanically scanable communication systems [4].

Analysis and design of LL and other lenses frequently involves a number of approximations. Optical methods [4] neglect both finite size of the lens and finite beam width of the source. In electromagnetic simulations, the source is commonly taken as a dipole [5], although in practice it is normally a horn. Therefore, taking the feed as a Huygens element [6] was an elegant step ahead. However, this source (crossed electric and magnetic dipoles) has the field of the fixed shape. Besides, it is essentially a 3-D source that cannot be used in the 2-D modeling. Even more adequate is taking the incident field as a Gaussian beam [7]. However, the latter as well as a series of the wave harmonics multiplied with a ‘window’ function, does not satisfy the Helmholtz equation. Hence, an error entering the solution is uncontrollable and may spoil the analysis of such a fine effect as WG.

A. V. Boriskin (e-mail: a_boriskin@yahoo.com) and A. I. Nosich are with the Institute of Radiophysics and Electronics, National Academy of Sciences, ul. Proskury 12, Kharkov 310085, Ukraine.

Home page: www.ire.kharkov.ua/dep12/MOCA

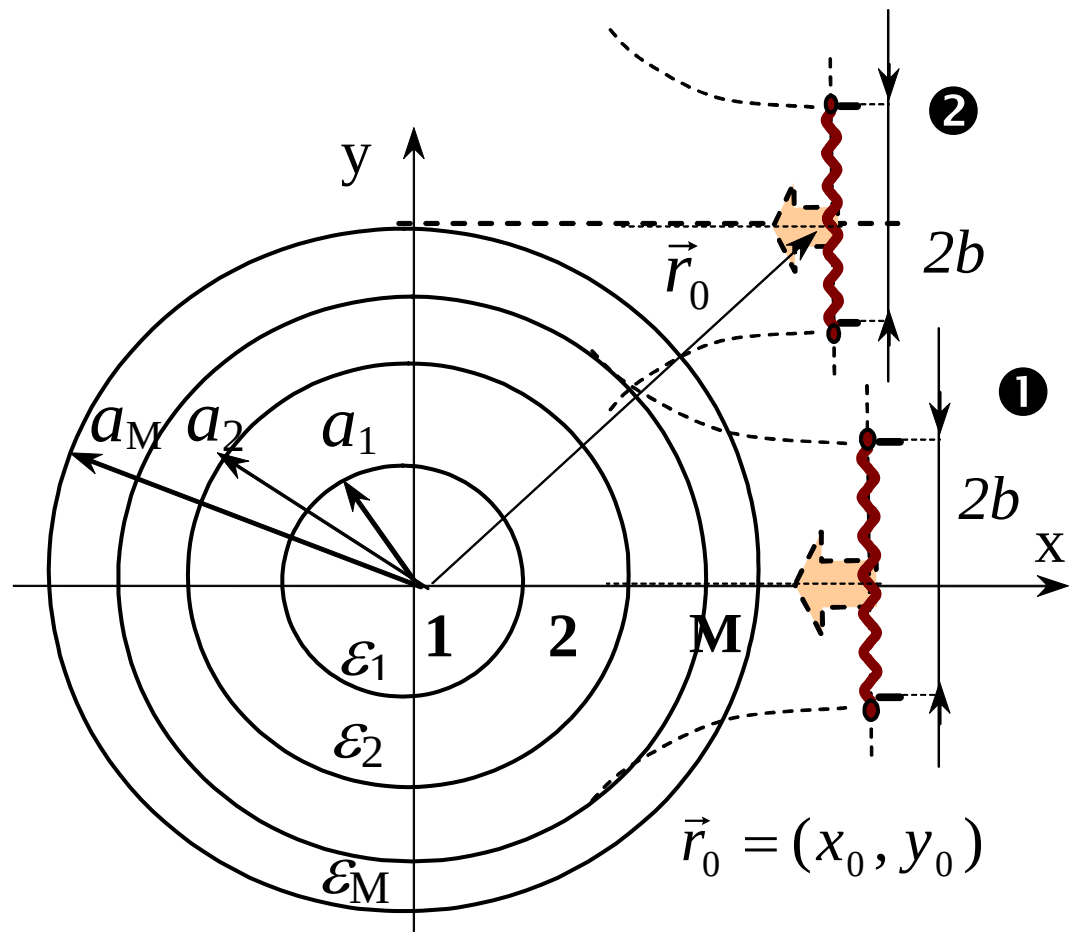

Fig.1. The scattering geometry of the frontal (1) and grazing (2) CSP illumination of a layered dielectric cylinder Dots and curvy lines denote branching points and branching cuts in the real space due to CSPs.

## II. Mathematical Model

The scattering geometry and notations are shown in Fig.1. An *M*-layer circular dielectric cylinder of the outer radius $a_M$ is illuminated by a time-harmonic ( $\sim exp(-i\omega t)$ ) beam generated by a CSP located at $\vec{r}_{cs}$. The main advantages of the CSP field

$$U_0 = H_0^{(1)}\left(k\left|\vec{r}-\vec{r}_{cs}\right|\right) \qquad (1)$$

are that it is an exact solution of the Helmholtz equation with respect to the observation point:

$$\left(\Delta_r + k^2\right)U_0(\vec{r},\vec{r}_{cs}) = 4i\delta\left(\vec{r}-\vec{r}_{cs}\right) \qquad (2)$$

and has a variable beam width controlled by parameter $kb$ [8], where $k$ is the free-space wavenumber, and the complex source position vector is

$$\vec{r}_{cs} = \left\{x_{cs}, y_{cs}\right\} = \vec{r}_0 + i\vec{b}\ ,\ \vec{r}_0 = \left\{x_0, y_0\right\},\ \vec{b} = \left\{b\cos\beta, b\sin\beta\right\} \qquad (3)$$

Parameters $b$ and $\beta$ are normally associated with the aperture width and orientation angle of a horn antenna simulated by a CSP. Note that the CSP excitations depicted in Fig. 1 imply that $\beta = \pi$ . Hence, the branching points of $U_0$ in the real space are located at $(x_0, y_0 \pm b)$.

Directive character of such the field (1) is well seen if the asymptotic is taken for $r \to \infty$ :

$$U_0 \approx \sqrt{2/i\pi kr} \cdot e^{ikr} e^{kb\cos(\varphi-\beta)} \tag{4}$$

Total field must satisfy the Helmholtz equation with the wavenumber $k_s = k\sqrt{\varepsilon_s}$ in the *s*-layer, the radiation condition at infinity, and the boundary conditions. The latter can be written as follows: if $r = a_s$

$$U_s = U_{s+1}, \qquad \frac{\partial U_s}{\partial r} = \alpha_{s+1}^2 \frac{\partial U_{s+1}}{\partial r} \tag{5}$$

where parameter $\alpha_s$ depends on the polarization: $\alpha_s = Z_s$ or $1/Z_s$ in the *E*- or *H*-case, respectively, and $Z_s = (\varepsilon_s/\mu_s)^{1/2}$ is the normalized impedance of the *s*-layer.

In the LL case, we shall assume the layers to be all of the same thickness $a_M/M$, with the relative dielectric constants taken in accordance with the LL rule [3,6]:

$$\varepsilon_s = 2 - (s - {}^1/_2)^2 / M^2 \tag{6}$$

Here and further index *s* runs from *1 to M* ( $\varepsilon_{M+1} = 1$).

Due to the axial symmetry of the scatterer, the total field in each layer can be written in terms of the azimuthal Fourier series [9]:

$$U_s^{tot} = \sum_{n=-\infty}^{\infty} \left[A_n^s J_n(k_s r) + B_n^s H_n(k_s r)\right] \cdot e^{-in\varphi} , \tag{7}$$

Here, $A_n^s$ and $B_n^s$ are the unknown coefficients; $H_n$ is the Hankel function of the 1st kind, and $J_n$ is the Bessel function. In the most inner layer, $B_n^1 = 0$ . Outside the cylinder $k_{M+1} = k$ , and the term with the $A_n^{M+1}$ coefficient corresponds to the incident field (1) translated to the cylinder coordinates by using the addition theorems for cylindrical functions:

$$U_0 = \begin{cases} \sum_{n=-\infty}^{\infty} J_n(kr_{cs}) H_n^{(1)}(kr) \cdot e^{-in\varphi_{cs}} e^{-in\varphi}; & r > |r_{cs}| \\ \sum_{n=-\infty}^{\infty} J_n(kr) H_n^{(1)}(kr_{cs}) \cdot e^{-in\varphi_{cs}} e^{-in\varphi}; & r < |r_{cs}| \end{cases}, \tag{8}$$

where

$$r_{cs} = \sqrt{r_0^2 - b^2 + 2ib(x_0\cos\beta + y_0\sin\beta)} \tag{9}$$

$$\varphi_{cs} = \arccos\left(\frac{x_0 + ib\cos\beta}{r_{cs}}\right) \tag{10}$$

The axial symmetry of the boundary conditions enables one to use them separately for each value of index *n* [6]. Here, as follows from the lower line of (8), outside the cylinder near its boundary,

$$A_n^{M+1} = H_n(kr_{cs}) \cdot e^{-in\varphi_{cs}} \tag{11}$$

On applying the boundary conditions sequentially, outward from the boundary between the two most inner layers, the following expression for the scattered-field expansion coefficient outside the cylinder is obtained:

$$B_n^{M+1} = \gamma_{Mn} H_n(kr_{cs}) \cdot e^{-in\varphi_{cs}} \tag{12}$$

where the coefficient $\gamma_{Mn}$ is determined by using the recurrent formulas:

$$\gamma_{0n} = 0, \tag{13}$$

$$\theta_{sn} = \theta_{sn}\left(\gamma_{(s-1)n}\right) = \frac{\gamma_{(s-1)n} H_n(k_s a_s) + J_n(k_s a_s)}{\alpha_s\left(\gamma_{(s-1)n} H_n'(k_s a_s) + J_n'(k_s a_s)\right)} \tag{14}$$

$$\gamma_{sn} = \gamma_{sn}(\theta_{sn}) = \frac{J_n(k_{s+1}a_s) - \theta_{sn}\alpha_{s+1} J_n'(k_{s+1}a_s)}{\theta_{sn}\alpha_{s+1} H_n'(k_{s+1}a_s) - H_n(k_{s+1}a_s)} \tag{15}$$

Out of the scatterer the total field can be finally written as a sum of the scattered and incident fields:

$$U_{M+1}^{tot} = U_{M+1}^{sc} + U_0 = \begin{cases} \sum_{n=-\infty}^{\infty} \Psi_n(kr_{cs}) H_n(kr) \cdot e^{-in\varphi}, \; r > |r_{cs}| \\ \sum_{n=-\infty}^{\infty} H_n(kr_{cs}) \cdot \Phi_n(kr) \cdot e^{-in(\varphi+\varphi_{cs})}, \; r < |r_{cs}| \end{cases}$$

where (16)

$$\Psi_n(kr_{cs}) = \left[\gamma_{Mn} H_n(kr_{cs}) + J_n(kr_{cs})\right] \cdot e^{-in\varphi_{cs}} \tag{17}$$

$$\Phi_n(kr) = \gamma_{Mn} H_n(kr) + J_n(kr) \tag{18}$$

Having determined the expansion coefficient outside the cylinder and applying the boundary conditions backward from the outer boundary, the following recurrent formulas for the other expansion coefficients are obtained:

$$A_n^s = \alpha_{sn}^a A_n^{s+1} + \beta_{sn}^a B_n^{s+1}, \; B_n^s = \alpha_{sn}^b A_n^{s+1} + \beta_{sn}^b B_n^{s+1} \tag{19}$$

where

$$\alpha_{sn}^{a,b} = \left\{J_n(k_{s+1}r_s) - \chi_{sn}^{a,b} J_n'(k_{s+1}r_s)\right\} / \zeta_{sn}^{a,b} \tag{20}$$

$$\beta_{sn}^{a,b} = \left\{H_n(k_{s+1}r_s) - \chi_{sn}^{a,b} H_n'(k_{s+1}r_s)\right\} / \zeta_{sn}^{a,b} \tag{21}$$

$$\zeta_{sn}^a = J_n(k_s r_s) - \chi_{sn}^a J_n'(k_s r_s) \tag{22}$$

$$\zeta_{sn}^b = H_n(k_s r_s) - \chi_{sn}^b H_n'(k_s r_s) \tag{23}$$

$$\chi_{sn}^a = H_n(k_s r_s) / \{\xi_{sn} H_n'(k_s r_s)\} \tag{24}$$

$$\chi_{sn}^b = J_n(k_s r_s) / \{\xi_{sn} J_n'(k_s r_s)\} \tag{25}$$

$$\xi_{sn} = k_s / k_{s+1} \tag{26}$$

Total radiated power is found by integrating the Poynting vector radial component over a circle of large radius $r \to \infty$. This enables one to replace the functions $H_n(kr)$ by their asymptotics, therefore

$$P_{rad} \equiv P_{M+1}^{tot} = \frac{\alpha_0 \pi}{2} \sum_{n=-\infty}^{\infty} \left|\Psi_n\right|^2 \tag{27}$$

where $\alpha_0$ is $Z_0$ or $Z_0^{-1}$, in the case of *E*- and *H*-polarization, respectively, $Z_0$ is the free-space impedance.

In the case of a lossy scatterer, the absorbed power can be calculated by integrating the same quantity over the outer surface of cylinder:

$$P_{abs}^{tot} = \frac{\alpha_0}{2k} \mathrm{Re} \sum_{n=-\infty}^{\infty} \Phi_n(ka_M) \frac{\partial \Phi_n^*(ka_M)}{\partial r} \cdot \left| H_n(kr_{cs}) \cdot e^{-in\varphi_{cs}} \right|^2 \tag{28}$$

In order to validate the results, the absorbed power was also calculated as an integral over the cross section, that was reduced to a series as well. Expression for this series is not given here because of its bulky form.

Based on the series representation, main-lobe directivity is determined as:

$$D = 2\pi \left| \sum_{n=-\infty}^{\infty} \Psi_n \cdot e^{-in\varphi_{\max}} \right|^2 \cdot \left( \sum_{n=-\infty}^{\infty} \left|\Psi_n\right|^2 \right)^{-1} \tag{29}$$

where $\varphi_{\max}$ is the main lobe direction.

Unlike a plane wave, CSP feed illuminates the scatterer in non-uniform manner. Therefore, it is natural to introduce some quantity for a reasonable characterization of this feature. We borrow this quantity from the theory of reflector antennas and call it *edge illumination* that is just the ratio of two values of the incident beam field, at the lens "edge" and at its center: $W=20 \log(U_0|_{edge}/U_0|_{center})$.

Numerical results shown below are related to the absorbed and radiated power fractions. They are normalized by the power radiated by a CSP located in the free space:

$$P_0 = \frac{\alpha_0 \pi}{2} \sum_{n=-\infty}^{\infty} \left| J_n(kr_{cs}) \cdot e^{-in\varphi_{cs}} \right|^2 = \frac{\alpha_0 \pi \cdot I_0(2kb)}{2} \tag{30}$$

where $I_0$ is the modified Bessel function.

## II. NUMERICAL RESULTS

A fast algorithm for the calculation of cylindrical functions with complex arguments, based on the recurrent technique (forward one for the Neumann and backward for the Bessel functions) was used [10]. These functions were computed with the machine precision.

Of course, for each *n*, the full set of all the boundary equations can be written and solved as a simultaneous matrix equation of the order *2M*. However, this requires greater computer resources and time than a recurrent scheme. It should be noted that the moment-method and FDTD approximations have been reported to suffer of heavy loss of accuracy near the high-Q natural frequencies [11]. Discussion of the nature of this defect can be found in [12], [13]. Therefore their application to study the narrow resonances such as WG ones has a limited value. The series solution is free of that demerit.

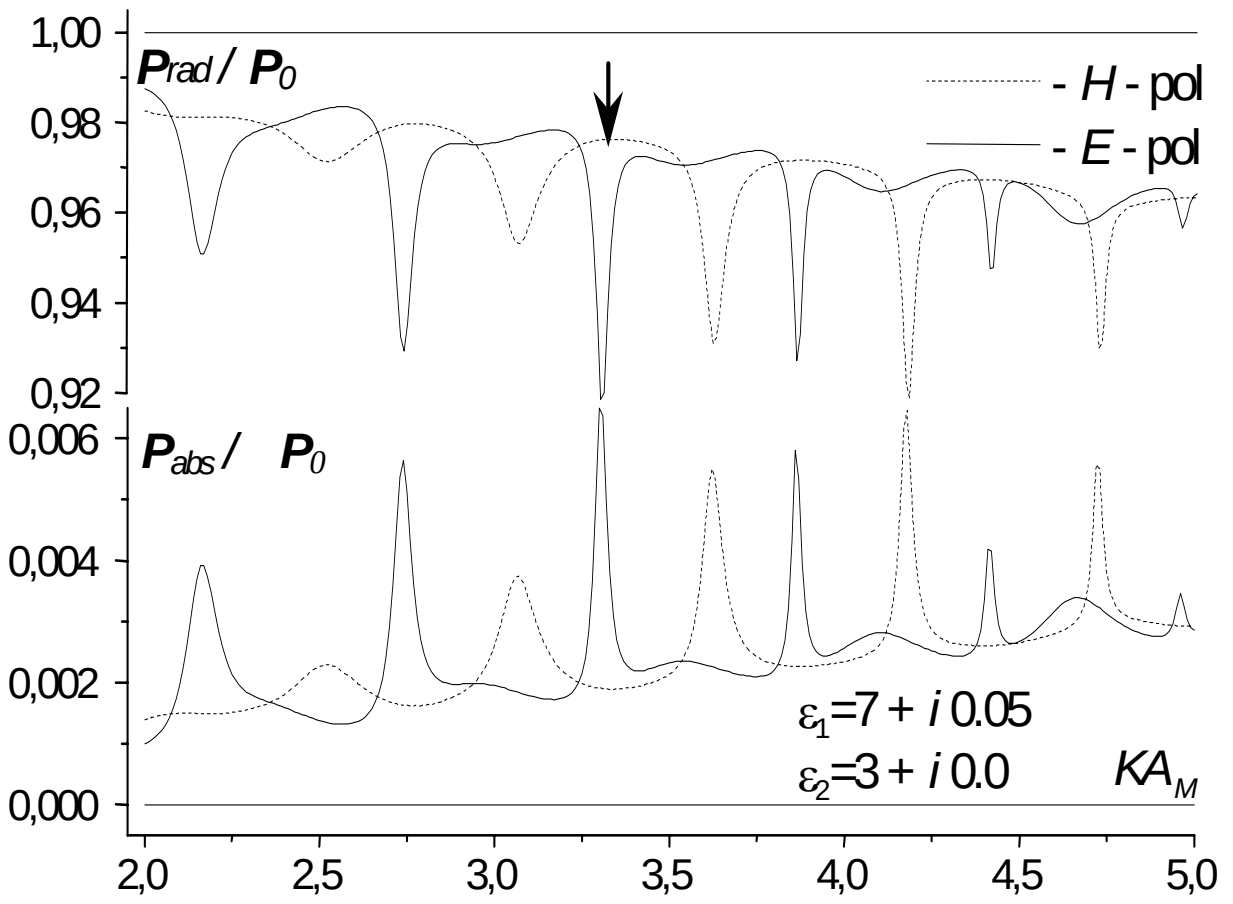

Fig. 2. Dissipated powers versus the normalized frequency ($M = 2$, $kb = 5$, $x_0/a_2 = 10$, $y_0/a_2 = 1$, $a_1 / a_2 = 0.8$).

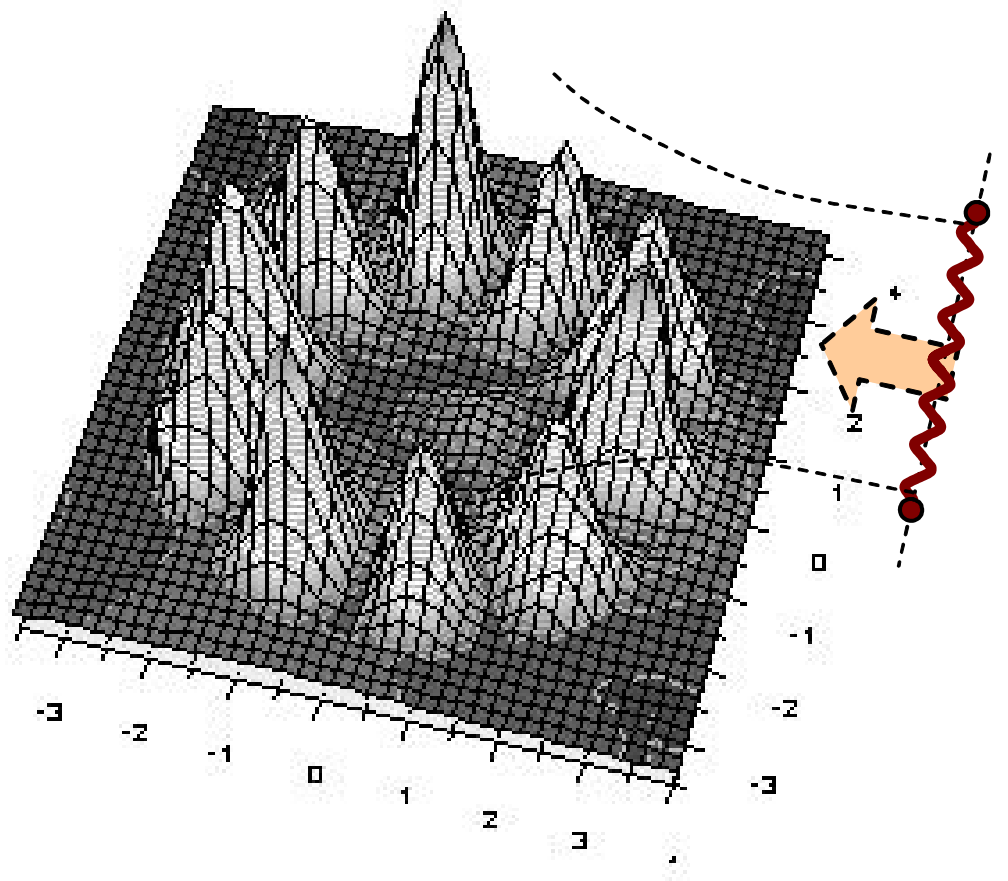

Fig.3. Near-field pattern of the $WGH_{5,1}$ resonance ($ka_M = 3.304$, marked by arrow in Fig.2).

In Figs. 2, presented are the frequency scans of the total radiated power and the absorbed power, in the grazing-beam excitation of a two-layer cylinder.

One can see periodic sequences of sharp resonances characteristic for the WG effect [1]. In a WG resonance, a greater part of the feed power is dissipated in the lossy material of the scatterer at the expense of a drop in the radiated power. Near-field portrait of a $WGH_{5,1}$ resonance is demonstrated in Fig.3.

The further numerical results relate to discrete LL modeling.

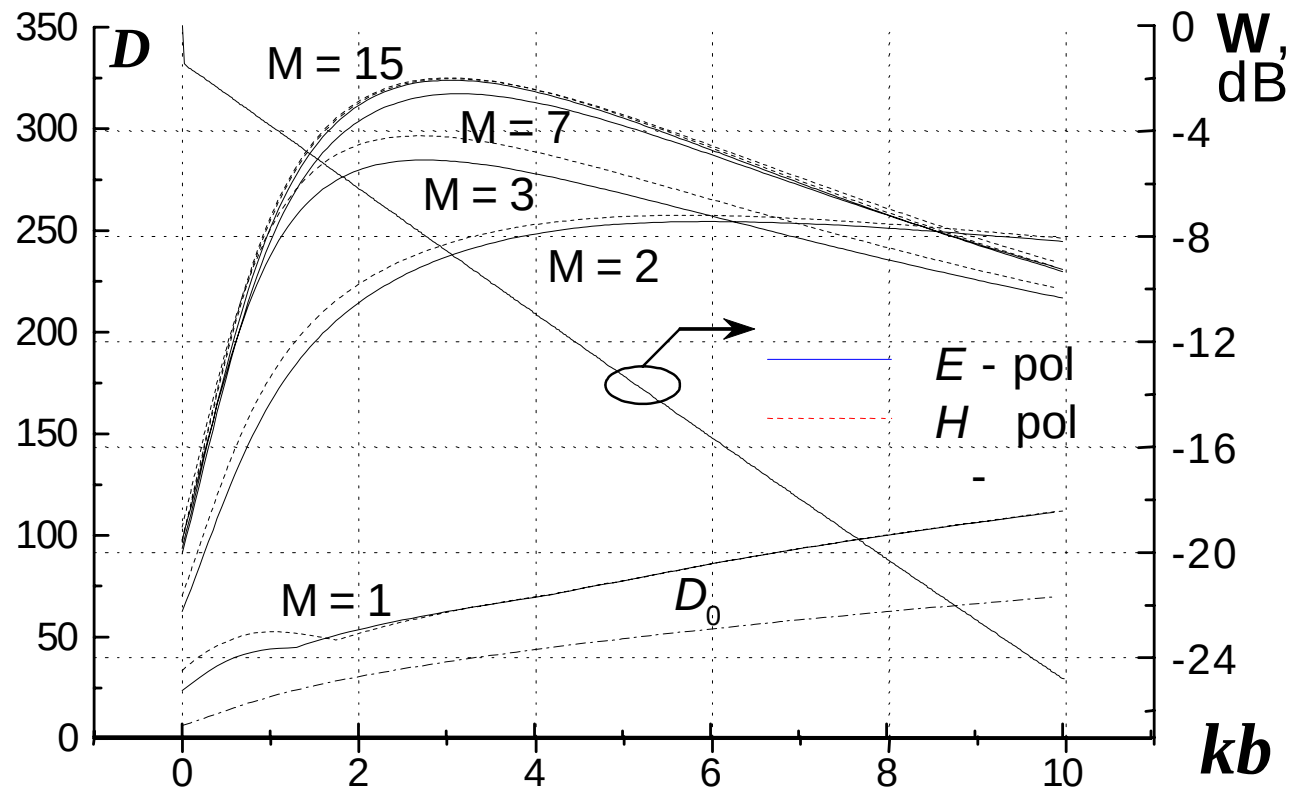

Fig.4. Directivity of LL and the lens edge elimination (shown by the arrow) versus the beam-width parameter ($ka_M = 60.0$, $x_0/a_M = 1.05$). $D_0$ is the directivity of CSP in the free space.

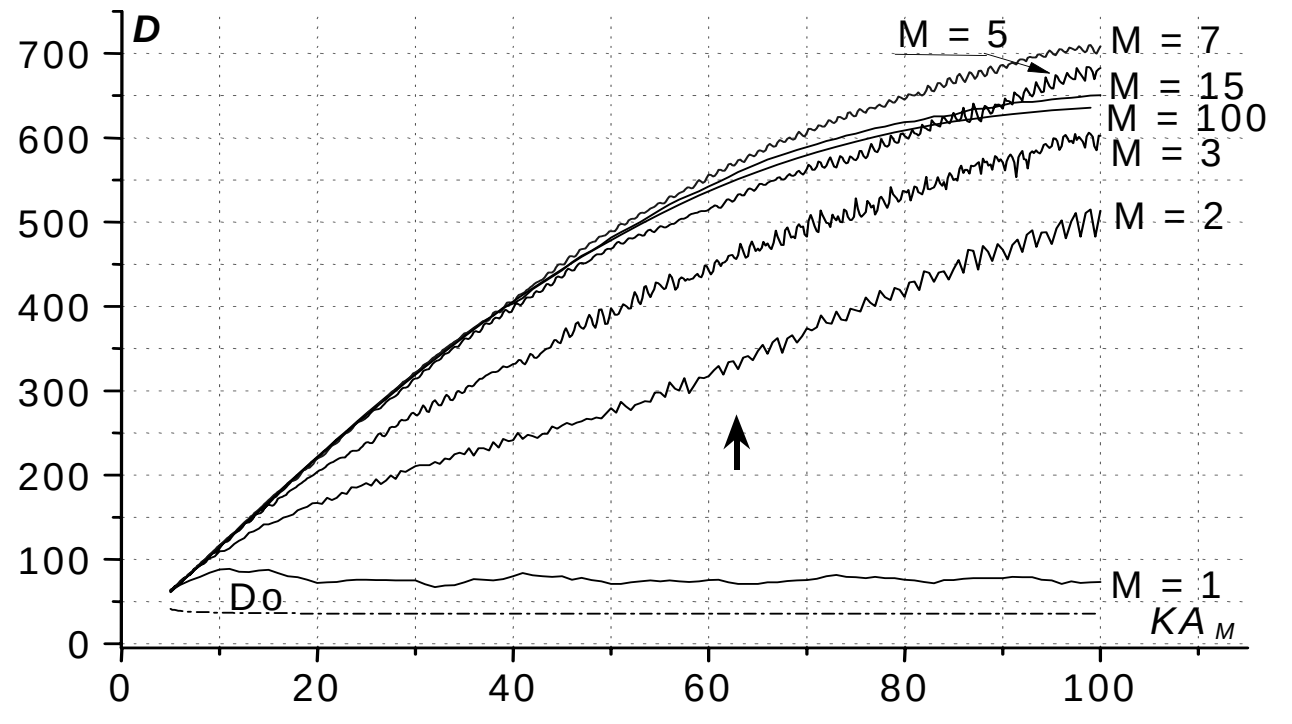

Fig.5. Directivity of LL versus the normalized frequency under the fixed –8dB edge illumination (*E*-pol, $x_0/a_M = 1.05$).

The family of curves in Fig. 4 enables one to optimize the LL performance in the case of the fixed geometry of the lens, by changing the edge illumination.

Here, a CSP source illuminates a multi-layered cylinder in symmetric manner and is placed near to its surface. This is because in the Geometrical Optics approximation the focus is just on the LL surface [3,6]. One can see that the minimum acceptable number of layers is two, although even a uniform cylinder displays focusing features; the optimal edge illumination is approximately –8 dB. Further study shows that the appropriate choice of the lens geometry, under a fixed edge illumination, can also improve LL performance. Note that WG are still excited but not so strongly as at the grazing beam incidence (Figs. 5 and 6). This effect is more visible if the number of layers is small.

In order to prove that the drops in the directivity as a function of frequency are explained by the WG mode excitation, the total filed in and near the lens has been calculated. A characteristic WG mode field portrait is well seen in Fig. 7. The fact that WG modes can be excited that results in sizable directivity loss proves a limited applicability of the geometrical and physical optics approaches in the case of the wavelength comparable with the lens size.

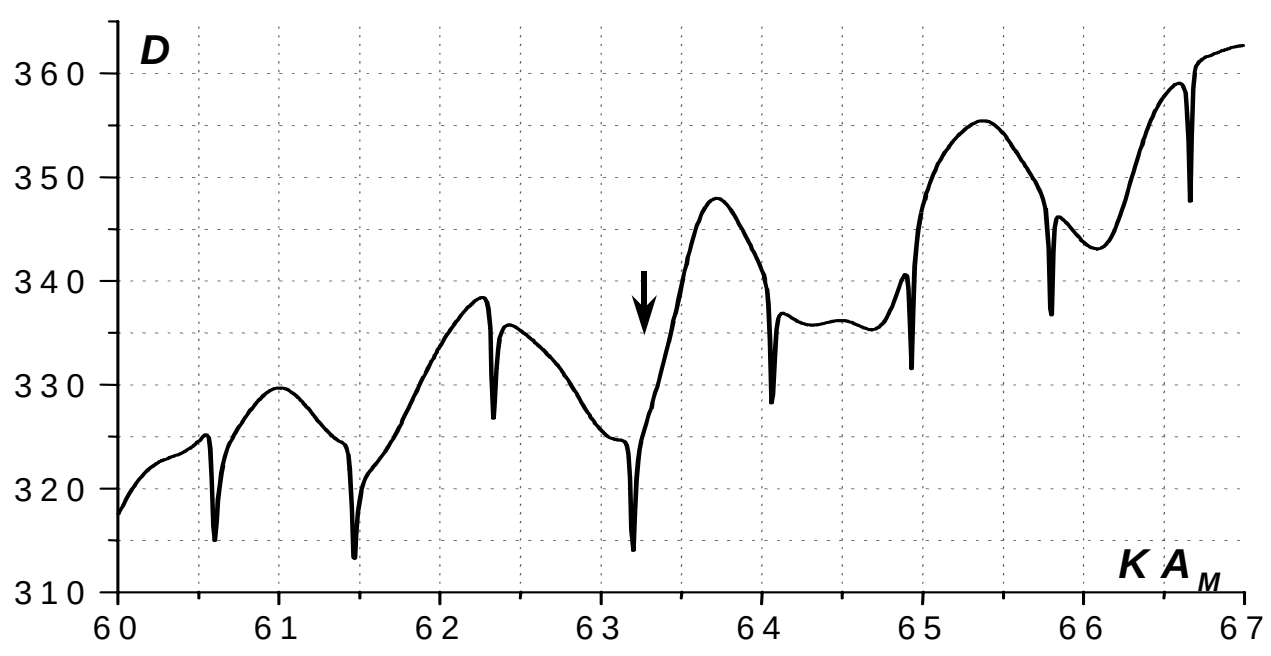

Fig. 6. Zoom of one of the curves in Fig. 5 (*E*-pol, $M = 2$, $\varepsilon_1 = 1.9375$, $\varepsilon_2 = 1.4375$).

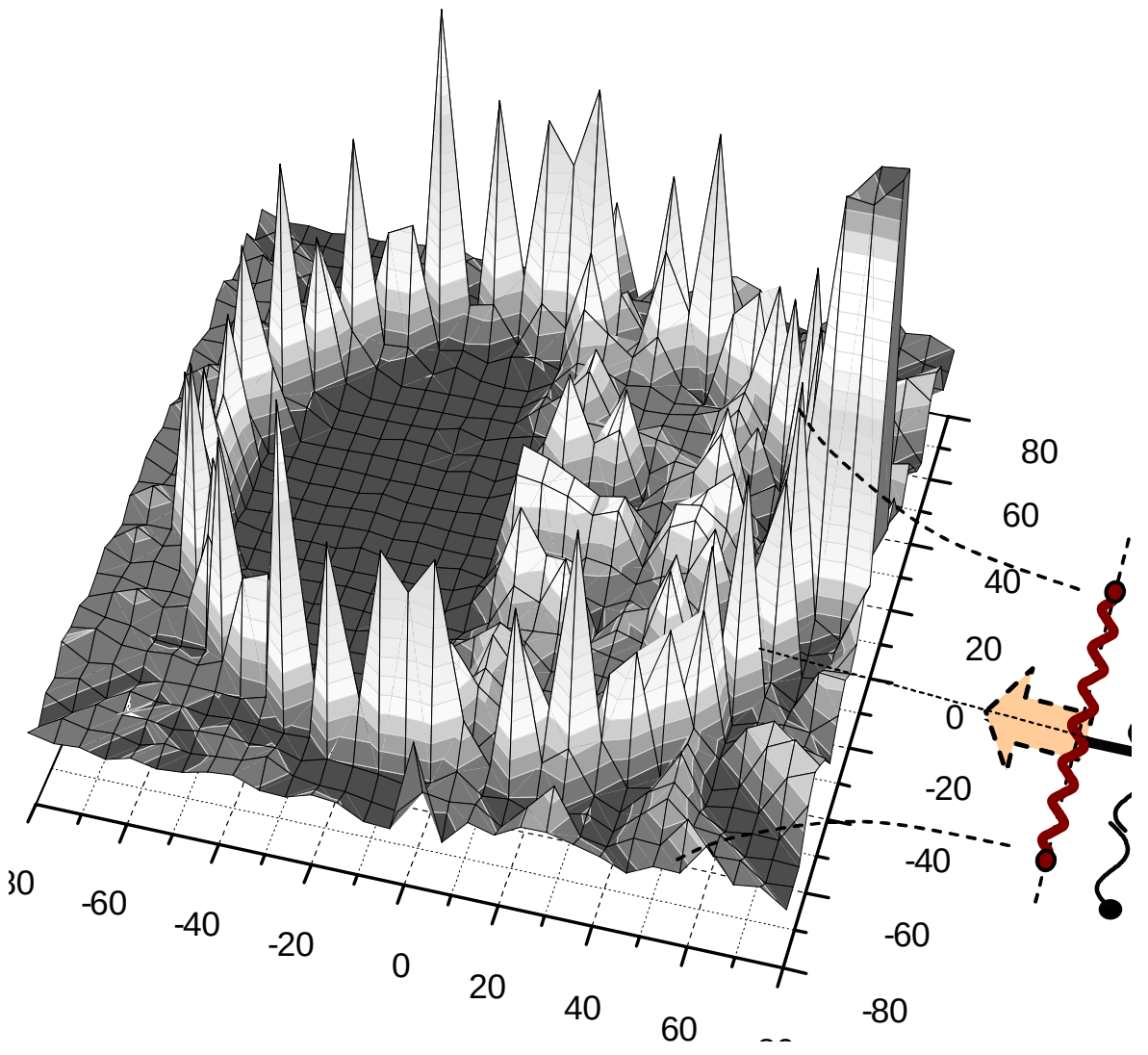

Fig.7. Near-field pattern of the $WGH_{16,1}$ resonance in LL ($kb = 2.76$, $ka_2 = 63.1973$, marked by arrows in Figs. 5, 6).

## III. Conclusions

Numerically exact series solution of a CSP-beam-fed multi-layer dielectric cylinder has been used to study the WG mode excitation and the LL effect. Accurate results have been presented for the both polarization cases and for the various cylinder and beam parameters. The analysis of LL excited by a CSP in frontal arrangement has shown that even a 3-layer 30-λ lens fed with the –8 dB edge illumination provides up to 17 fold improvement in directivity. It has been demonstrated that the WG modes are excited by the CSP field in grazing illumination as well as in frontal mode. They are accompanied by the drops in the LL directivity and scattered power, and the peaks in the absorbed one. The depths of the drops depend on the number of layers: the lager the number of layers the smaller the drops. Thus, the results, presented in the paper, complete the previous analyses done by approximate approaches with a study of fine wavelength-scale effects in discrete LL.

## IV. Acknowledgment

The authors are grateful to S. V. Boriskina for the aid in computations and to the reviewers for helpful advises. The first author was supported with the IEEE MTT Society Graduate Fellowship Award in Microwave Engineering.

## V. References

[1] J. R. Wait, "Electromagnetic whispering gallery modes in a dielectric rod," *Radio Science*, vol. 2, no. 9, pp. 1005-1007, 1967.

[2] S. V. Boriskina, A. I. Nosich, "Radiation and absorption losses in a whispering-gallery-mode dielectric resonator excited by a dielectric waveguide," *IEEE Trans. Microwave Theory Techniques*, vol. MTT-47, no. 2, pp. 234-241, Feb. 1999.

[3] R. K. Luneburg, *The Mathematical Theory of Optics*, Brown University Press, 1941.

[4] C. A. Fernandes, "Shaped dielectric lenses for wireless millimeter-wave communications," *IEEE Antennas Propag. Mag*., vol. AP-41, no. 5, pp. 141-151, Oct. 1999.

[5] A. T. Greenwood, J.-M. Jin, "A field picture of wave propagation in inhomogeneous dielectric lenses," *IEEE Antennas Propag. Mag., v*ol. AP-41, no.5, pp. 9-18, Oct. 1999.

[6] H. Mieras, "Radiation pattern computation of a spherical lens using Mie series," *IEEE Trans. Antennas Propagat.*, vol. AP-30, no.6, pp. 1221-1224, Nov. 1982.

[7] E. E. M. Khaled, S. C. Hill, P. W. Barber, "Scattered and internal intensity of a sphere illuminated with a Gaussian beam," *IEEE Trans. Antennas Propagat.*, vol. 4, no. 3, pp. 1221-1224, Mar. 1999

[8] L. B. Felsen, "Complex-point source solutions of the field equations and their relation to the propagation and scattering of the Gaussian beams," *Symp. Mathem.*, vol. 18, pp. 39-56, 1975.

[9] J. R. Wait, *Electromagnetic Radiation from Cylindrical Structures*, NY: Pergamon Press, 1959.

[10] C.F. Du Toit, "The numerical computation of Bessel functions of the first and second kind for integer orders and complex arguments," *IEEE Trans. Antenna Propagat*., vol.38, no.9, pp.1341-1349, 1990.

[11] G. L. Hower, R. G. Olsen, D .J. Earls, J. B. Schneider, "Inaccuracies in numerical calculation of scattering near natural frequencies of penetrable objects," *IEEE Trans. Antennas Propag.*, vol. AP-41, no. 7, pp. 982-986, July 1993.

[12] D. G. Dudley, "Error minimization and convergence in numerical methods," *Electromagnetics*, vol. 5, no. 2-3, pp. 89-97, 1985.

[13] S. D. Senturia, N. Aluru, J. White, "Simulating the behavior of micro-electro-mechanical system devices: computational methods and needs," *IEEE J. Comput. Science Engineering*, vol. 4, no. 1, pp. 30-43, 1997.

**Artem V. Boriskin** (S'99) was born in Kharkov, Ukraine, in 1977. He received his M.S. degree in Radio-Physics from the Kharkov State University, in 1999.

He is currently a Ph.D. student at the Institute of Radio-Physics and Electronics of the National Academy of Sciences of Ukraine.

His research interests are in the EM scattering by various dielectric objects with applications to dielectric lenses, rod antennas and antenna radomes.

He received an IEEE Microwave Theory and Technique Society Graduate Student Fellowship Award in Microwave Engineering and a TUBITAK (Scientific and Technical Research Council of Turkey) - NATO Fellowship Award in 2000 and 2001, respectively.

**Alexander I. Nosich** (M'94, SM'95) was born in 1953 in Kharkov, Ukraine. He had earned his M.S., Ph.D. and D.Sc. degrees, all in Radio-Physics, in 1975, 1979 and 1990, respectively, from the Kharkov State University.

Since 1979, he is on research staff of the Institute of Radio-Physics and Electronics of the National Academy of Sciences of Ukraine, in the departments of electronics and computational electromagnetics. In 1992-2000, he held visiting professorships and guest-scientist fellowships in Turkey, Japan, Italy and France. In 1995, he was the organizer and first chairman of the IEEE AP-S East Ukraine Chapter, the first one in the Former Soviet Union (now joint with AES, ED, EMB, GRS, MTT and NPS Societies ). Since 1995, he is on the editorial board of the Microwave and Optical Technology Letters. In 1990-2000, he has been an organizer and technical program committee co-chairman of the series of international conferences on Mathematical Methods in EM Theory (MMET) in USSR and Ukraine.

His research interests include wave scattering, radiation, propagation and absorption studied by the analytical regularization techniques, and the history of microwaves.